\documentclass[twocolumn]{aastex63}
\usepackage{amsmath,amstext,natbib}
\usepackage{color,soul}
 
\accepted{by ApJ on Jan, 2022}

\shorttitle{Transiting Star-sized Dust Clump in HD 166191}
\shortauthors{K.Y.L. Su et al.}

\begin{document}

\title{A Star-sized Impact-produced Dust Clump in the Terrestrial Zone of the HD 166191 System}

\author[0000-0002-3532-5580]{Kate Y.~L.~Su}
\affiliation{Steward Observatory, University of Arizona, 933 N Cherry Avenue, Tucson, AZ 85721--0065, USA}

\author[0000-0001-6831-7547 ]{Grant M. Kennedy}
\affiliation{Department of Physics, University of Warwick
, Gibbet Hill Road, Coventry CV4 7AL, UK}
\affiliation{Centre for Exoplanets and Habitability, University of Warwick, Gibbet Hill Road, Coventry CV4 7AL, UK}

\author[0000-0001-8291-6490 ]{Everett Schlawin}
\affiliation{Steward Observatory, University of Arizona, 933 N Cherry Avenue, Tucson, AZ 85721--0065, USA}

\author[0000-0003-4393-9520]{Alan P. Jackson}
\affiliation{School of Earth and Space Exploration, Arizona State University, 550 E. Tyler Mall, Tempe, Arizona, 85287, USA }

\author[0000-0003-2303-6519]{G. H. Rieke}
\affiliation{Steward Observatory, University of Arizona, 933 N Cherry Avenue, Tucson, AZ 85721--0065, USA}
\affiliation{Lunar and Planetary Laboratory, The University of Arizona, 1629 E. University Boulevard, Tucson, AZ 85721--0065, USA}

\correspondingauthor{Kate Su}
\email{ksu@as.arizona.edu}

\begin{abstract}

We report on five years of 3--5 $\mu$m photometry measurements obtained by warm {\it Spitzer} to track the dust debris emission in the terrestrial zone of HD 166191 in combination with simultaneous optical data.  We show that the debris production in this young ($\sim$10 Myr) system increased significantly in early 2018 and reached a record high  level (almost double by mid 2019) by the end of the {\it Spitzer} mission (early 2020), suggesting intense collisional activity in its terrestrial zone likely due to either initial assembling of terrestrial planets through giant impacts or dynamical shake-up from unseen planet-mass objects or recent planet migration. This intense activity is further highlighted by detecting a star-size dust clump, passing in front of the star, in the midst of its infrared brightening. We constrain the minimum size and mass of the clump using multiwavelength transit profiles and conclude that the dust clump is most likely created by a large impact involving objects of several hundred kilometers in size with an apparent period of 142 days (i.e., 0.62 au assuming a circular orbit). The system's evolutionary state (right after the dispersal of its gas-rich disk) makes it extremely valuable to learn about the process of terrestrial-planet formation and planetary architecture through future observations. 

\end{abstract}


\section{Introduction} 
\label{sec:intro} 

Theoretical simulations indicate that the very existence of terrestrial planets depends on the collisional merging of planetary embryos and oligarchs \citep{asphaug98,agnor99,asphaug06_hit_and_run} over a period of $\sim$200 Myr after protoplanetary disks have cleared (e.g., \citealt{chambers13,raymond14}). Copious lines of evidence suggest that such giant impacts were common and played a central role in terrestrial-planet formation in the early solar system \citep{wyatt_jackson16}. An efficient way to probe this process around other stars is through the infrared-excess emission of the dust produced in these impacts. Theoretical calculations for the general time evolution of mid-infrared excesses during this period \citep{kenyon04b,kenyon16} are qualitatively consistent with the observations showing that the amounts of mid-infrared excesses decay with time \citep{rieke05,su06,meng17}, and that each of the giant impacts injects new debris into a system's terrestrial zone, creating stochastic variations in the mid-infrared outputs of young stars that are actively forming terrestrial planets \citep{kenyon05,genda15}.   

However, an inconvenient truth is that the terrestrial-planet formation interpretation of these young, extremely dusty systems is not unique. An alternative possibility is that the high dust levels result from the transient dynamical clearing of regions inhabited by planetesimals. In the solar system, such a process must have happened when the asteroid belt's Kirkwood gaps were initially cleared by Jupiter. Following the dispersal of the gas disk, which stabilizes orbits against perturbations, the orbital excitation of asteroids in Jupiter's resonances would have led to increased collision rates and dust production, and a brief period of extreme dustiness. The V488 Per system might be the result of an analogous process -- an intense planetesimal excitation by nearby planetary or substellar companions \citep{rieke21_v488per}. 

Models predict that the observable signatures of giant impacts during the terrestrial-planet formation period would persist over millions of years as mid-infrared excesses \citep{kenyon04b,jackson12,kral15}.
Although {\it Spitzer} has found some such dusty debris systems around young stars (see a recent review by \citealt{chen_su_xu20}), the detection rate is low -- a few percent to 10\% depending on the observing wavelengths and ages of the samples \citep{currie07_h_chi_per,balog09,carpenter09_feps,kennedy_wyatt13,meng17}. This appears to be in stark contrast with the high incident rate of small (0.5--1.5 $R_{\oplus}$) planets around solar-like stars inferred from {\it Kepler}  \citep[e.g.,][]{bryson21_eta_earth_Kepler}. Several mechanisms have been proposed to explain the apparent discrepancy, such as that the large planetary scale collisions might not be as frequent as models predict and that the impact-produced material has a shorter observable lifetime due to gas drag or anisotropic nature of escaping debris  \citep{kenyon_najita_bromley2016,watt21}.

Furthermore, we have also witnessed dramatic changes in infrared output associated with individual planetesimal collisions. Multiyear warm {\it Spitzer} monitoring programs have shown that disk variability in these extreme debris disks provides diagnostics of conditions in terrestrial-planet formation by measuring collisional outcomes in the time domain. These observations measure collisional outcomes and extract the physical properties of the impact-produced orbiting dust clumps as seen in the prototype of extreme systems, NGC 2354--ID8 \citep{su19}. Impact-produced dust clumps have been inferred from either detecting quasi-periodic infrared modulations \citep{meng14,su19} or irregular optical dipping events  \citep{dewit13_rzpsc,gaidos19_hd240779,powell21_tic400799224}. Optically thick dust clumps can also shield tiny grains from radiation pressure blowout and allow them to accumulate as they are produced from larger bodies through collisional cascades  or as impact-produced vapor condenses \citep{johnson12b}. The stable mid-infrared solid-state features in HD 172555 and HD 113766A (two young dusty debris disks), over roughly two decades, further corroborate the existence of impact-produced dust clumps in the extreme debris disks \citep{su20}. 

Here we report on five years of warm {\it Spitzer} monitoring observations of HD 166191, accompanied by simultaneous ground-based optical data.  HD 166191 is a young late F- to early G-type star at a distance of 101 pc, surrounded by a large amount of circumstellar dust as identified by its infrared excess \citep{oudmaijer92,clarke05,kennedy_wyatt13,fujiwara13,schneider13,kennedy14}. 
Although the structure of the HD 166191 disk has not been resolved by any prior observations, it is complex as inferred by the spectral energy distribution (SED), showing the presence of (1) hot dust traced by 3--5 $\mu$m emission, (2) prominent solid-state features in the mid-infrared indicative of small, warm dust, and (3) cold dust traced by far-infrared and mm observations \citep{schneider13,kennedy14,garcia_hughes_hd166}.

New infrared and optical observations are described in Section \ref{sec:sec2} where we show that the HD 166191 system experienced a significant increase (almost double) in the 3--5 $\mu$m flux in early 2018 and reached a plateau in mid 2019. On top of the flux increase, a dip in the light curves was observed at both infrared and optical wavelengths. We also review and summarize the system's basic properties like age, mass, and luminosity that we adopt for further interpretation. In Section \ref{sec:transits_comp}, we derive the properties of the multiwavelength dip and a prior dip observed in the optical using simple functional forms and a physical (curtain) model and characterize the obscuring object in terms of size and orbital location. In Section \ref{sec:discussion}, we first discuss the cause of the observed variability by assessing the properties of the infrared emission between the quiescent and active states (Section \ref{sec:general_intepretation}), discuss the properties of the obscuring object (Section \ref{sec:clump_evo}), and conclude that a giant collision between two large ($\gtrsim$ Vesta size) bodies is the most likely explanation (Section \ref{sec:implication}). We also briefly discuss the implication of nondetection of the short-term flux modulation expected from an impact-produced debris clump (Section \ref{sec:short-term}), and speculate on the trigger mechanisms of the two quiescent and active states (Section \ref{sec:trigger}). 
A conclusion is given in Section \ref{sec:conclusion}.

\section{Observations and Results} 
\label{sec:sec2}

\begin{figure*}
    \includegraphics[width=\linewidth]{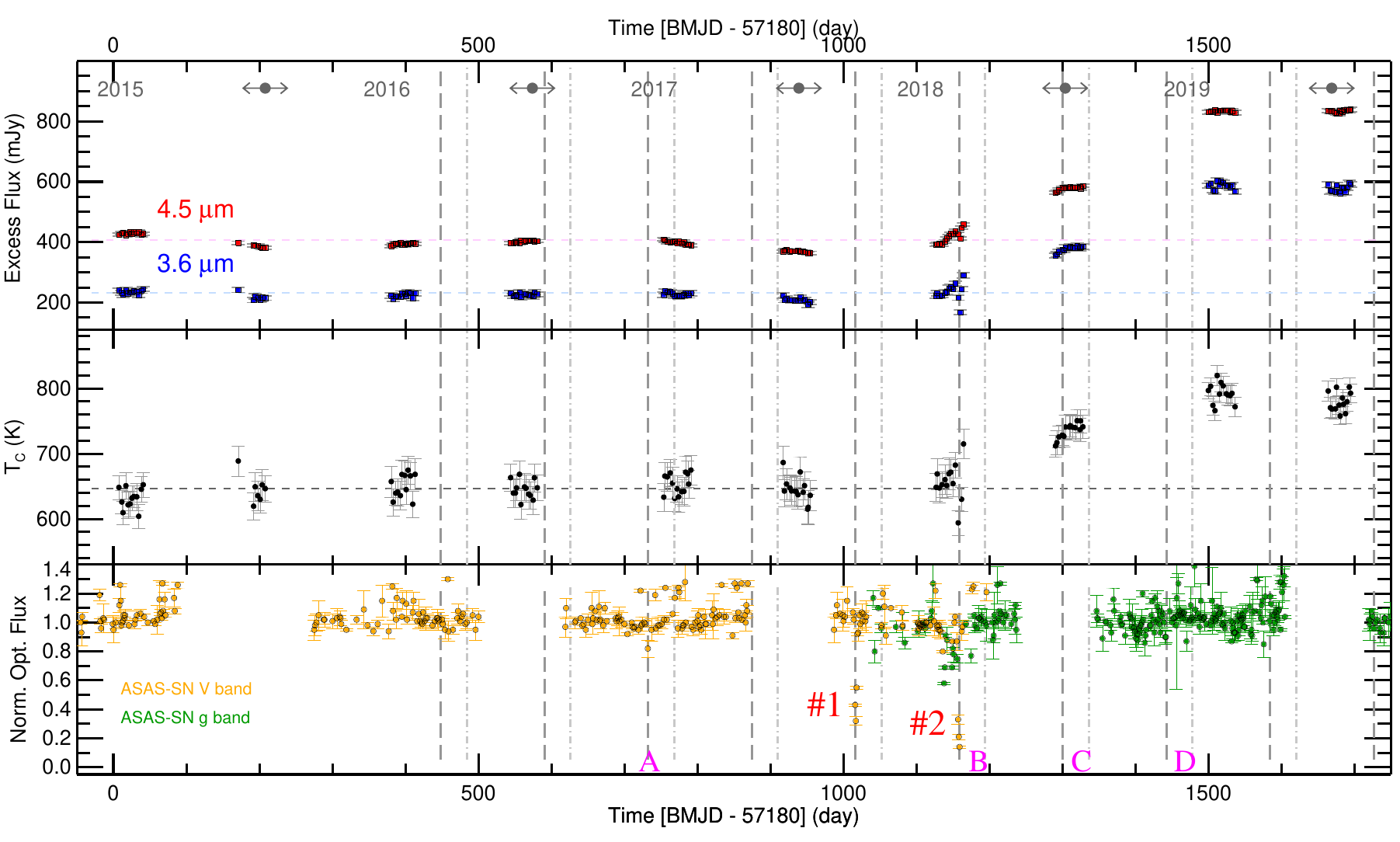}
    \caption{Time-series observations for the HD 166191 system where the upper two panels show the data obtained with warm {\it Spitzer} expressed as the amount of excess emission at 3.6 (blue) and 4.5 (red) $\mu$m and the corresponding color temperature ($T_c$) estimated by the flux ratio, and the bottom panel shows the normalized optical ($V$ and $g$ bands) photometry obtained by the ASAS-SN project. All error bars are shown as 1$\sigma$. The horizontal dashed lines in the upper two panels represent the median values of the disk fluxes and temperatures. The vertical dashed lines in the bottom panel mark the interval of 142 days defined by the two deep dips in the optical data (see the upper panel of Figure \ref{fig:opt_irac_transit} for a zoom-in view). The second deep dip was also observed in the {\it Spitzer} data (lower panels of Figure \ref{fig:opt_irac_transit}). The dashed-dotted lines mark the times when the dust clump passes the disk ansae (i.e., an offset of one-forth of the orbital period; for details, see the discussion in Section \ref{sec:short-term}.)  }
    \label{fig:irac_hd166191}
\end{figure*}

\subsection{Warm {\it Spitzer}/IRAC} \label{sec:irac_obs}

Warm {\it Spitzer} observations were obtained under GO programs PID 11093,  13014, and 14266 (PI Su) with data covering from 2015 June to the end of the Spitzer mission in 2020 January. HD 166191 has two {\it Spitzer} visibility windows per year, each a total of $\sim$39 days in length. During each visibility window, a cadence of 3$\pm$1 days was used to monitor the system, which was designed to search for any variation due to material as close as 0.1 au from the star. In 2015 November, six planned observations were lost due to an unexpected safe mode in the {\it Spitzer} operation. A total of 126 sets of observations at both 3.6 and 4.5 $\mu$m bands were obtained. We used a frame time of 0.4 s with 10 cycling dithers (i.e., 10 frames per Astronomical Observation Request (AOR)) to minimize the intra-pixel sensitivity variations of the detector \citep{reach05_irac} at both bands, achieving a signal-to-noise ratio of $\gtrsim$170 in single-frame photometry. These data were first processed with IRAC pipeline S19.2.0 by the {\it Spitzer} Science Center.  We performed aperture photometry on each of the data sets, following the procedure outlined in \citet{su19}. The final weighted-average photometry is given in Table \ref{tab:irac}. We also determined the instrumental repeatability by monitoring several isolated sources in the field of view. A $\sim$1\% rms in the measured photometry was found for nonvarying sources in these data.  

Using Kurucz stellar atmospheric models (details see Section \ref{sec:star}), we estimated the stellar photosphere in the two IRAC bands to be 483 and 315 mJy with a typical uncertainty of 1.5\% limited by the accuracy of the optical and near-infrared photometry. Assuming the star is stable at these wavelengths, the infrared excesses (i.e., the disk fluxes) and associated color temperatures are shown in Figure \ref{fig:irac_hd166191} over the five-year span. Overall, the infrared output of the system at 3.6 and 4.5 $\mu$m has two states: a quiescent phase before 2018 and an active phase afterward. During the quiescent state, the disk emission is relatively stable: 223$\pm$11 mJy and 397$\pm$19 mJy at 3.6 and 4.5 $\mu$m, respectively, resulting in a color temperature of 645$\pm$20 K. The disk has become brighter since early 2018, reaching a maximum level by mid 2019 (587$\pm$12 mJy and 835$\pm$4 mJy at 3.6 and 4.5 $\mu$m, respectively), and has been relatively stable since, with a higher color temperature of 790$\pm$20 K. In the early active state, a rapid flux drop (eclipse) occurred near the Barycentric Modified Julian Date (BMJD) of 58340 (label B in the bottom of Figure \ref{fig:irac_hd166191}) at both wavelengths. During the eclipse, the color temperature also dropped, reaching a minimum of $\sim$600 K, which is consistent with less heating from the star due to obscuration. This eclipse event is further discussed in Section \ref{sec:transits_comp}.

\subsection{Optical ASAS-SN Data} \label{sec:opt-asas-sn}

Optical data were extracted from the ASAS-SN Sky Patrol (\url{https://asas-sn.osu.edu/}, \citealt{shappee14, kochanek17}). These publicly available optical data were taken in the $V$- and $g$-band filters using a number of different cameras. We first normalized the photometry to the average value with the data obtained in the same camera and filter from 2015 to 2020.   After normalizing, we binned the data for each of the bands by averaging to a cadence of 1 day. The normalized optical light curves are shown in the bottom panel of Figure \ref{fig:irac_hd166191}. The typical rms is about 5--10\% determined by all $\sim$700--800 measurements in each of the filters, which is $\sim$2--3 times noisier than that of the stars in the same field as HD 166191. The noisy nature is consistent with the star being a young solar-like star, with the expected high level of chromospheric activity \citep[e.g.,][]{lockwood2007}. Despite the noisy nature, two sharp flux drops, separated by $\sim$142 days, were found in these optical data. The upper panel of Figure \ref{fig:opt_irac_transit} shows a zoom-in view of this region. We verified the transit signals are authentic by examining the ASAS-SN data of TYC 6843-1878-1 (a star 2\farcm6 away) and derived a photometry rms of 3.5\%, suggesting any dimming more than 10\% of the nominal value might be significant; however, we only consider dips that are more than 30\% ($\gtrsim$3$\sigma$), given the noisy nature of the star. The two sharp drops show an eclipse depth of more than 70\% and appear to be relatively symmetric in the ingress and egress sides. Hereafter, we refer to the first event as dip \#1 and the second as dip \#2, which is also observed in the warm {\it Spitzer} data (lower panels of Figure \ref{fig:opt_irac_transit}). Detecting dip \#2 in two different telescopes clearly establishes that it is associated with the star, not other nonastronomical phenomena. As guided by the separation of $\sim$142 days, we searched for other possible dimming events before and after. No deep (more than 30\%) events were found in the ASAS-SN light curves before and after the two deep events, although there are some complex, tentative dipping events in the $\sim$10--20\% levels. We first focus on deriving the basic properties of the deep dips \# 1 and \# 2 in Section \ref{sec:transits_comp}, and will assess the nature of other tentative events in Section \ref{sec:clump_evo}.

\subsection{Optical HAO Data}
\label{sec:haodata}

To further discover and characterize potential dips at later times, we obtained eight epochs of optical photometry of the system from 2019 May 11 to 2019 May 24, 2019 (covering display dates of 1434--1447 in Figure \ref{fig:irac_hd166191}) at the Hereford Arizona Observatory (HAO) (\url{http://www.brucegary.net/HAO/}). We observed with the HAO AstroTech 0.41 m Ritchey-Chretien telescope equipped with a Santa Barbara Instrument Group ST-10XME CCD camera and with an $r'$-band filter. Image sets with exposure times of 2 s were obtained. Standard bias, dark, and flat-field calibrations were performed before photometry measurements were made relative to a single reference star with an $r'$ magnitude from the APASS catalog \citep{henden16_apass_catalog}, resulting in a typical photometry uncertainty per measurement (epoch) of 0.015--0.020 mag. We found no optical variability in this two-week window with an average mag of 8.556$\pm$0.015 mag. We further discuss these data in Section \ref{sec:clump_evo}. 

\begin{table}
\begin{center}
\tablewidth{0pc}
\caption{Basic parameters and references for HD 166191 adopted in This paper for the distance, stellar temperature ($T_{\ast}$), radius ($R_{\ast}$), luminosity ($L_{\ast}$), mass ($M_{\ast}$), age, measured radial velocity ($V_r$), and rotation velocity ($v sini$).   \label{tab:stellar}}
\begin{tabular}{lcl}
\hline
\hline
Parameter  &  Value  & Reference \\
\hline 

$L_{\ast}$  & 4.1 $L_{\sun}$  & 1 \\
$R_{\ast}$  & 2 $R_{\sun}$  & 1 \\
$T_{\ast}$  & 6000 K  & 2, 3, 4 \\
$M_{\ast}$  & 1.6 $M_{\sun}$  & 1 \\
Distance & 101.2$\pm$0.2 pc & 5 \\
Age  &  $\sim$10 Myr  & 1, 4 \\
$V_r$  & $-$8.5$\pm$1.5 km s$^{-1}$  & 2, 3, 4 \\
$v sin i $  & 27 $\pm$1 & 4 \\
\hline
\end{tabular}
\end{center}
\tablenotetext{}{References: [1] this work; [2] \citet{schneider13}; [3] \citet{kennedy14}; [4] \citet{potravnov18}; [5] {\it Gaia} EDR3. } 
\end{table}

\subsection{System Properties}
\label{sec:star}

The properties of the central star in the HD 166191 system are reasonably well constrained. A number of high-resolution optical spectra \citep{schneider13,kennedy14,potravnov18} together indicate an effective temperature of 6000 K or slightly higher. The lack of optical and near infrared emission lines and optical spectroscopic variability suggest no active gas accretion onto the star. The age of the star has been muddled by its apparent association with the $\sim$5 Myr-old Herbig Ae star HD 163296, but the situation has been clarified by \citet{potravnov18}. This reference identified a group of five other stars that share similar kinematics and youth indicators ($\sim$10 Myr) to HD 166191 and suggested they belong to the extended Corona Australis Association \citep{CrA_2008ref}. The radial velocities ($-8.1 \pm 1.3$ km s$^{-1}$ from \citet{schneider13}, $-10.1 \pm 1$ km s$^{-1}$ from  \citet{kennedy14}, and $-7.4 \pm 2$ km s$^{-1}$ from  \citet{potravnov18}) indicate no close companion, and there is no indication of binarity on the sky.  \citet{potravnov18} estimate a rotation velocity $v\text{sin}i=$ 27$\pm$1 km~s$^{-1}$. Given that typical F5 main-sequence stars have rotation velocities of $\sim$30 km~s$^{-1}$, the stellar rotation axis is then roughly perpendicular to the line of sight. The extinction to HD 166191 is negligible \citep{ruiz2018}. {\it Gaia} gives an accurate distance of 101.2$\pm$0.24 pc (Gaia EDR3, \citealt{gaia16,gaia_edr3}), replacing the much lower accuracy value of 119 pc from Hipparcos, which many prior studies adopted. Although the stellar activity results in a modest level of variability, long-term monitoring does not detect it as a large-amplitude variable \citep{otero2020}. Given the accurate distance, the stellar luminosity is estimated to be 4.1 $L_\odot$ by integrating the stellar atmospheric models. 

A summary of the intrinsic properties is that the star is at a $T_\text{eff} \sim 6100$ K, 10 Myr old, and has a luminosity of $\sim$4.1 $L_\odot$. We have interpreted these characteristics in terms of stellar evolution using two sets of isochrones: the classic ones by \citet{siess2000} and the ones from the Padova/Trieste group, described by \citet{bressan2012}. A consistent, if not perfect, fit is provided for ages of 8--11 Myr, a mass of $\sim$1.6 $M_\odot$, and a radius of $\sim$2.0 $R_\odot$. Note that the nominal stellar radius for a late F-type main-sequence star is $\sim$1.4 $R_{\sun}$, and the enlarged stellar size is consistent with the youth of HD 166191.
We used Kurucz atmospheric models with the derived parameters to estimate the stellar contribution in the two IRAC bands. The basic parameters that we adopt in this paper are summarized in Table \ref{tab:stellar}. 

The circumstellar disk responsible for the infrared emission of the system has been described as an extreme debris disk with an infrared fractional luminosity ($f_d = L_{\text{IR}}/ L_{\ast}$) of $\sim$0.1 \citep{schneider13} or as a transitional disk \citep{kennedy14}. It does not match either category perfectly. At the age of the star, the  majority of protoplanetary disks/transitional disks have dissipated but some protoplanetary/transitional ones persist \citep{balog2016,meng17}. However, the huge excess at $\sim$20 $\mu$m, {\it Spitzer} [8]--[24] = 2.9 or {\it WISE} W2--W4 = 4.7, is in the range expected of a protoplanetary or transitional disk. The colors [3.6]--[5.6] = 1.2 or W1--W2 = 0.8 are substantially too red for a traditional transitional disk (where [3.6]--[5.6] $<$ 0.4 is expected for transitional disks of similar age; \citep{balog2016}). This indicates that there is more material close to the star than in a true transitional system. The prominent solid-state feature in the 10 $\mu$m region is similar to many extreme debris disks such as HD 113766A (also an F-type star), where the disk is known to consist of two separate inner ($\sim$500 K) and outer ($\sim$150 K) components \citep{olofsson13,su20}. Because the infrared-excess emission in HD 166191 can be well described by two distinct dust temperatures: $\sim$760 and $\sim$175 K \citep{schneider13}, it is very likely that the disk in HD 166191 is two-belt-like. Finally, a definitive argument against the typical transitional-disk hypothesis is the lack of cold CO gas measured in the millimeter \citep{garcia_hughes_hd166}. Nonetheless, the mix of properties does suggest a unique system that illustrates a point in the evolution of a planetary system from the protoplanetary stage to a more mature state.

\begin{figure*}
    \includegraphics[width=\linewidth]{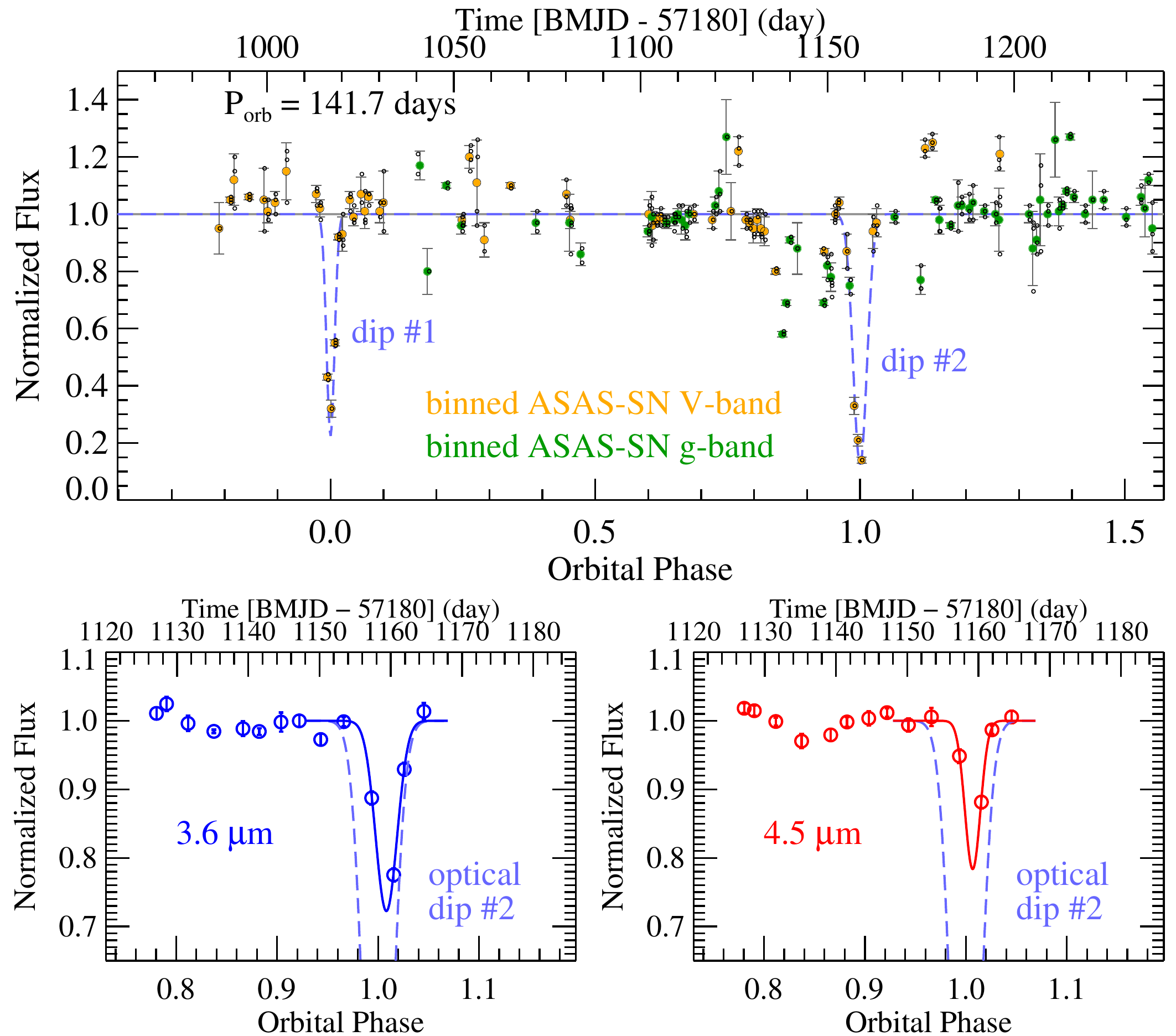}
    \caption{The upper panel shows the optical light curves centered on the two deep dips revealed by the ASAS-SN data. Data were folded with an orbital period of 142 days and initialized at the first dip. Small open circles depict the raw, unbinned data while the large filled circles with error bars show the 1 day binned data. Dashed blue curves are the curtain fits (details see Section \ref{sec:curtain}). The lower panels show the normalized infrared light curves for dip \#2 by assuming the occultation only blocks the star. The solid color lines are the curtain fits at 3.6 $\mu$m (blue) and 4.5 $\mu$m (red), while the the light-blue dashed line is the optical curtain model shown on the top.  The occulting object is less thick (both in depth and width) at longer wavelengths, and the center appears to be shifted later by $\sim$1 day (for details, see Section \ref{sec:curtain}). }
    \label{fig:opt_irac_transit}
\end{figure*}

\section{Multiwavelength Transit Model Comparison}
\label{sec:transits_comp}

\subsection{Functional Forms}
\label{sec:ahs_shs}

Among the two optical light curves, no flat-bottomed behavior (i.e., the area of the obscuring object is smaller than the cross section of the stellar disk) nor obvious asymmetry is seen with the 1 day sampling rate. Similarly, no asymmetry nor flat bottom is found in the {\it Spitzer} transit profiles given the coarse cadence of $\sim$3 days. We first used simple functional forms (a Gaussian profile or a symmetric hyperbolic secant \citep{rappaport14_koi2700b}) to quantify the basic properties of the transits such as the width and depth.  Unlike in the optical, the total infrared flux of the system increased over time during dip \#2 (Figure \ref{fig:irac_hd166191}), likely due to the increase of dusty debris. For the normalization of the infrared light curves, a linear function was included in the functional fits. We then flattened the infrared light curves by assuming two different scenarios: that the obscuring object occults the light from both the star and dusty disk or the star only. In terms of the transit depth between 3.6 and 4.5 $\mu$m, there is no significant difference between the two scenarios (blocking the star-only and star-and-disk cases) except that the depth is deeper ($\sim$2 times) when only blocking the star. In terms of the transit duration, on the contrary, there is a stark contrast between the two wavelengths -- the duration is longer at 3.6 $\mu$m ($\sim$2 times) compared to the one at 4.5 $\mu$m (Figure \ref{fig:opt_irac_transit}), and the trend is consistent between the star-only and star-and-disk scenarios. Including the functional fit of the dip \#2 at the $V$ band, the wavelength-dependent trend strengthens -- the transit depth gets shallower as the observed wavelength increases while the transit duration shows an opposite trend. 

Using the two $V$-band transit profiles, we further explored whether there is significant evolution between dips \#1 and \#2 by assuming they were caused by the same obscuring object. This assessment is complicated by the fact that the star is intrinsically noisy at optical wavelengths; a proper comparison should take the stellar intrinsic and systematic noises into account. We employed a Bayesian fit with a Gaussian dip function (amplitude and width) and a Gaussian Process (GP)\footnote{The Simple Harmonic Oscillator term was used for the Gaussian kernel.} regression to allow for correlated light-curve errors. The fits were performed in both the raw (i.e., unbinned) and the normalized, binned data. For the raw data, the GP reveals a typical flux uncertainty of $\sim$53 mJy, which is $\sim$10 times higher than the quoted raw flux uncertainty, but similar to the normalized and binned data (3.5\%; see Section \ref{sec:opt-asas-sn}.)  In both raw and binned datasets, we found that the Gaussian dip amplitude between the two dips is within 1$\sigma$ but the width is different by 2$\sigma$. This exercise suggests that (1) the normalized and binned data do properly capture the uncertainty of the optical light curve, and (2) there is apparent evolution between dips \#1 and \#2 at the $\sim$2$\sigma$ levels. Given the fact that there were no similar deep dips before and after the observed ones, it seems very likely that there is a rapid evolution in the obscuring object if the dips were caused by the same phenomenon. We will further explore the evolutionary nature under such an assumption and extract the basic properties of the obscuring object next.

\begin{table}
\begin{center}
\tablewidth{0pc}
\caption{Model parameters for the Curtain Model\tablenotemark{a}, Peak Optical Depth  $\tau$, FWHM $w$ (in $R_*$), Curtain Velocity $v$ (in stellar radii per day, $R_* d^{-1}$), Time When the Curtain Center Passes the Stellar Center $t_0$, and Delay between the Maximum Visible Transit and Maximum Infrared Transit $\Delta t$.  Subscripts $V$, 3.6, and 4.5, refer to data in the visible, 3.6~$\mu$m, and 4.5~$\mu$m bands respectively.   \label{tab:transitparameters}}
\vspace{-0.2cm}
\begin{tabular}{cccc}
\hline
\hline
Dip		& \#1	  &  \multicolumn{2}{c}{\#2}   \\
\hline 
\multicolumn{4}{l}{\it Fitting individual band} \\
$\tau_V$	& 0.79$^{+0.02}_{-0.02}$	  & \multicolumn{2}{c}{0.88$^{+0.01}_{-0.01}$}  \\
$w_V$		& 2.63$^{+0.09}_{-0.10}$	  & \multicolumn{2}{c}{4.57$^{+0.25}_{-0.25}$}  \\
$v_V$		& 2.50$^{+0.07}_{-0.07}$	  & \multicolumn{2}{c}{2.50$^{+0.07}_{-0.07}$}  \\
$t_{0,V}$	& 58197.00$^{+0.02}_{-0.02}$&\multicolumn{2}{c}{58338.73$^{+0.07}_{-0.07}$} \\
\hline
		& 	  &     Star only        &   Star + Disk   	\\
\cline{3-4}
$\tau_{3.6}$	& \nodata & 0.29$^{+0.02}_{-0.02}$ & 0.18$^{+0.01}_{-0.01}$ \\
$w_{3.6}$	& \nodata & 3.75$^{+0.23}_{-0.22}$ & 3.74$^{+0.22}_{-0.23}$ \\
$v_{3.6}$	& \nodata & 2.50$^{+0.07}_{-0.07}$ & {2.50$^{+0.07}_{-0.07}$}  \\
$t_{0,3.6}$	& \nodata & 58339.38$^{+0.07}_{-0.06}$&{58339.37$^{+0.07}_{-0.07}$} \\
\vspace{0.3cm}
$\tau_{4.5}$	& \nodata & 0.23$^{+0.19}_{-0.05}$ & 0.09$^{+0.05}_{-0.02}$ \\
$w_{4.5}$	& \nodata & 2.68$^{+0.35}_{-0.62}$ & 2.72$^{+0.34}_{-0.52}$ \\
$v_{4.5}$	& \nodata & 2.51$^{+0.07}_{-0.07}$ & {2.51$^{+0.06}_{-0.07}$}  \\
$t_{0,4.5}$	& \nodata & 58339.17$^{+0.11}_{-0.11}$&{58339.18$^{+0.10}_{-0.12}$} \\
\hline
\hline
\multicolumn{4}{l}{\it Fitting all three bands together with time offset} \\
		& 	  &     Star only        &   Star + Disk   	\\
\cline{3-4}
$\tau_V$  & \nodata & 0.88$^{+0.01}_{-0.01}$ & 0.88$^{+0.01}_{-0.01}$ \\
$w_V$		& \nodata & 4.59$^{+0.24}_{-0.26}$ & 4.58$^{+0.24}_{-0.24}$  \\
$\tau_{3.6}$ & \nodata & 0.29$^{+0.02}_{-0.02}$ & 0.19$^{+0.01}_{-0.01}$ \\
$w_{3.6}$	& \nodata & 3.71$^{+0.25}_{-0.22}$ & 3.70$^{+0.24}_{-0.21}$ \\
$\tau_{4.5}$ & \nodata & 0.19$^{+0.03}_{-0.02}$ & 0.08$^{+0.01}_{-0.01}$ \\
$w_{4.5}$	& \nodata & 2.88$^{+0.25}_{-0.23}$ & 2.91$^{+0.24}_{-0.25}$ \\
$v$		& \nodata & 2.50$^{+0.01}_{-0.01}$ & 2.50$^{+0.01}_{-0.01}$  \\
$t_0$	& \nodata & 58338.73$^{+0.07}_{-0.07}$ & 58338.72$^{+0.07}_{-0.06}$ \\
$\Delta t$ & \nodata &  0.59$^{+0.08}_{-0.09}$ & 0.60$^{+0.09}_{-0.08}$ \\
\hline
\end{tabular}
\end{center}
\tablenotetext{a}{Assuming a circular orbit with the velocity search range of 2.4--2.6 $R_{\ast} \text{day}^{-1}$.}
\end{table}

\subsection{Curtain Model}
\label{sec:curtain} 

The variable nature of the occultations argues that the transiting object is not a single opaque body such as a stellar companion.  Symmetric profiles at all wavelengths between the ingress and egress also suggest that the transiting material is symmetric along the direction of motion and likely lies at similar radii when transiting the star. The wavelength-dependent depths further suggest the transiting material has an optical depth gradient along the transit direction. To tie the multiwavelength transits together, we then adopt a simple one-dimensional, semiopaque curtain model, as described by \citet{kennedy17_rzpsc}, to fit the transit profiles. The curtain is assumed to have a Gaussian-like profile for the optical depth along the direction of motion, with a peak optical depth, $\tau$ and a width $w$ (FWHM) in units of stellar radius ($R_{\ast}$). The curtain moves at a constant velocity, $v$, in units of stellar radius per day ($R_{\ast}~ \text{day}^{-1}$) across the stellar disk, and the center of the curtain coincides with the stellar center at time $t_0$. As discussed in \citet{kennedy14}, the size and the velocity of the curtain are completely degenerate for a specific transit profile, i.e., the larger the curtain, the faster the resulting velocity. Under the assumption that dips \#1 and \#2 have the same origin (i.e., the orbital period is $\sim$142 days), we restrict the range of velocity search under the assumption that the curtain is on a circular orbit. In other words, the velocity is expected to be close to the Keplerian velocity (i.e., $\sim$47 km~s$^{-1}$ around a 1.6 $M_{\sun}$ star, or at $\sim$2.5 $R_{\ast}~ \text{day}^{-1}$ for $R_{\ast}\sim$2 $R_{\sun}$). A narrow range of velocities (2.4--2.6  $R_{\ast} \text{day}^{-1}$) was used to constrain the other three parameters. Much larger velocities, resulting in larger curtains, would be possible if the curtain is on an eccentric orbit. Unfortunately, we cannot constrain the curtain width under such a condition by only fitting the transit profiles, so we did not model these cases.

We determined the best-fit parameters and the associated uncertainties by matching the transit profile at each wavelength (binned $V,$ 3.6 and 4.5 $\mu$m data) first using the Python package $emcee$ \citep{emcee}. The derived parameters are summarized in Table \ref{tab:transitparameters}. By comparing the two optical dips (\#1 and 2), the simple curtain model suggests that the optical depth and the curtain width got thicker and wider with time. For dip \#2, both the optical depths and widths are much less in the infrared wavelengths compared to the ones in the optical, and there is an apparent shift in $t_0$ where the infrared transits occur later by $\sim$1 day.

\begin{figure}
\centering
    \includegraphics[width=0.5\textwidth]{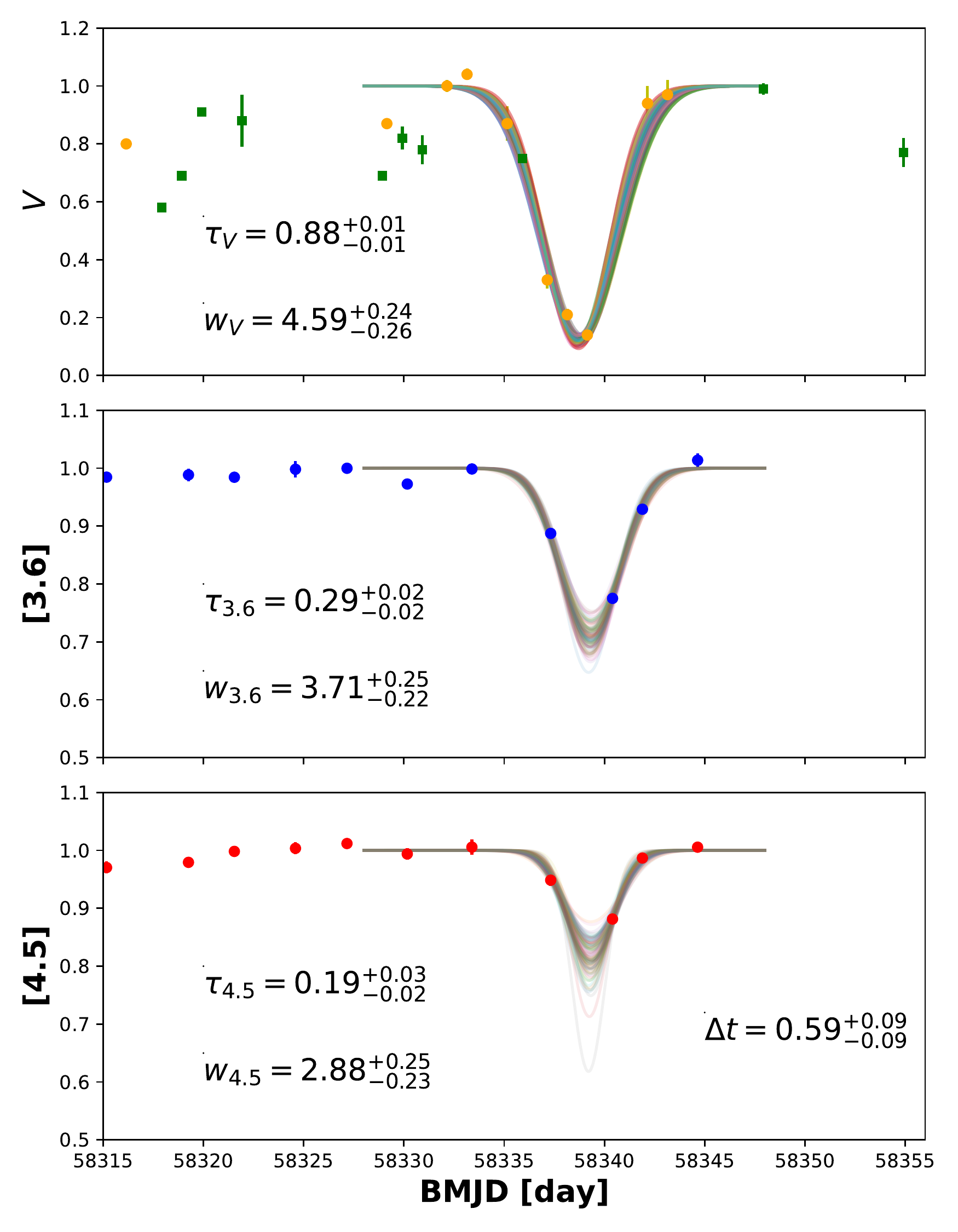}
    \caption{Flattened light curves for the dip \#2 with one of the best-fit curtain models on a circular orbit by fitting all three bands simultaneously. The color symbols are the same as Figure \ref{fig:opt_irac_transit}. The curtain is assumed to transit the star only. }
    \label{fig:curtainmodel}
\end{figure}

Because dip \#2 was detected at all three wavelengths, the fit was also performed by combining all three bands of data. To better quantify the shift, we also include a time difference ($\Delta t$) between the optical and infrared transits in the three-band data fit. The derived parameters are also shown in Table \ref{tab:transitparameters}, and Figure \ref{fig:curtainmodel} illustrates one of the fits for dip \#2. Overall, there is not much difference in the derived parameters either by fitting individual bands separately or together -- both indicate that the optical depth and the width of the curtain for dip \#2 were reduced significantly at longer wavelengths, and this trend remained the same whether the transiting object blocked the star only or both the star and the disk. Based on these characteristics, the obscuring object is most likely a clump of dust. The fairly symmetric shape between the ingress and egress suggests that particle sizes in the dust clump are either not significantly affected by radiation pressure within two orbits or the initial velocity distribution of the particles is more or less along the radial direction so that the alteration due to radiation pressure is minimal.

\begin{figure}
    \centering
    \includegraphics[width=\linewidth]{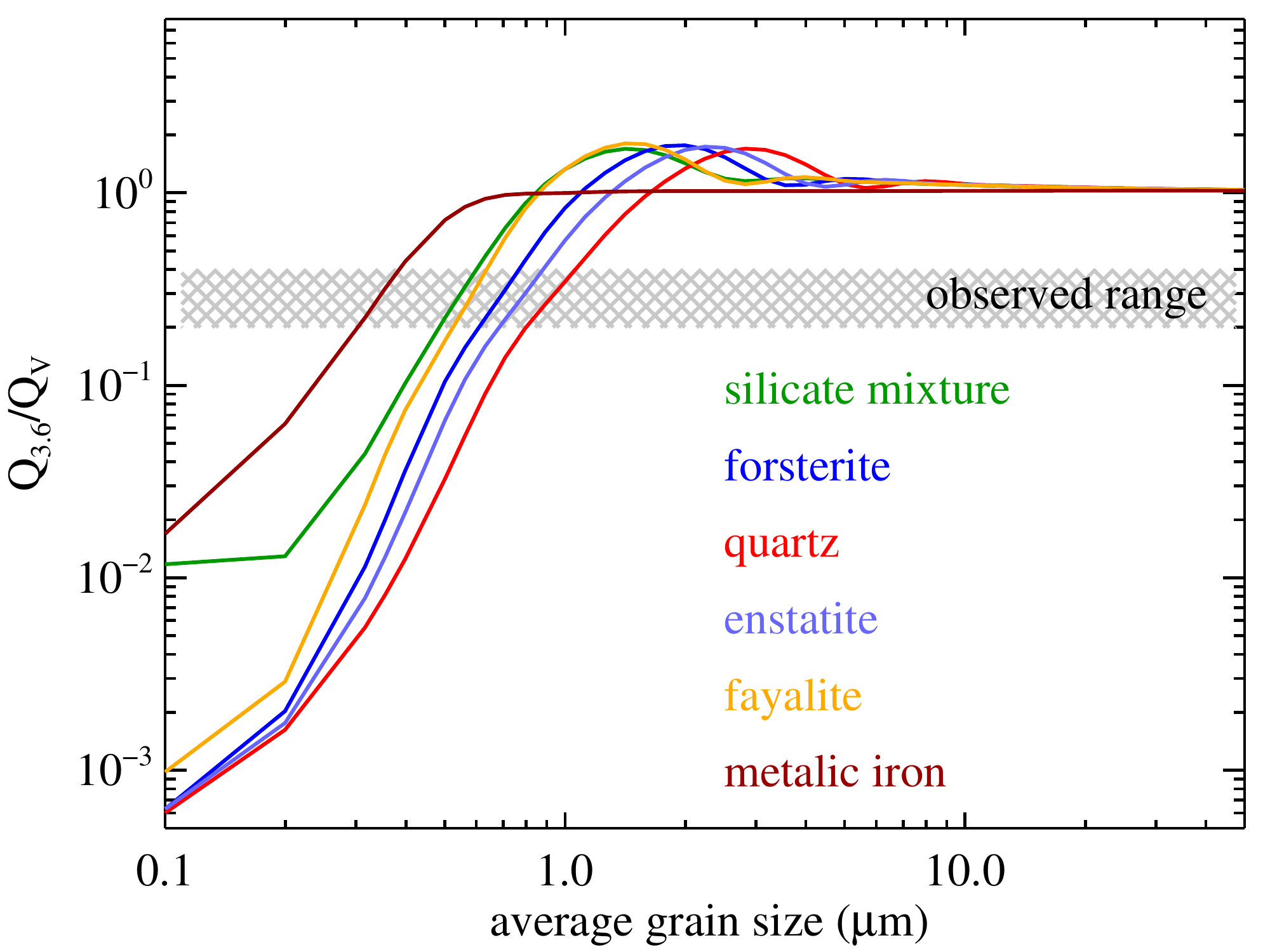}
    \caption{Extinction ratio ($y$-axis) between the $V$ and 3.6 $\mu$m bands for various compositions for dust grains in a log-normal size distribution where the average size is shown on the x-axis. The observed range for the dip \#2 dust clump is marked as the hashed region.}
    \label{fig:qratio}
\end{figure}

\section{Discussion} 
\label{sec:discussion}

The year-long, large infrared brightening seen in the warm {\it Spitzer} data, the detection of the deep dips in both the optical and infrared during the rapid phase of the infrared flux increase, combined with the fast evolution of the dip, all suggest that we are witnessing a large asteroid collision in the terrestrial zone of the HD 166191 system. Although we think the asteroid collision scenario best describes what is observed in HD 166191, it is necessary to point out that the deep dips might be caused by other independent means that are not related to the infrared brightening. For example, gas-related clumping and hydrodynamic instabilities in a gas-rich protoplanetary disk could produce irregular sharp dips \citep{bouvier03,dullemond03_uxo,bredall20_dippers_asas-sn} and would not be related to the brightening in the system's infrared output. However, the lack of abundant gas argues against such mechanisms operating in the HD 166191 system.  Here, we only focus on the asteroid collision hypothesis and discuss its implications. 

\subsection{What Causes the Observed Variability?}
\label{sec:general_intepretation}

Large-scale collisions among planetesimals and planetary embryos in the terrestrial-planet region are expected to be common after the dissipation of the gas in a protoplanetary disk \citep{chambers_wetherill98}, a stage we think HD 166191 is in. Such collisions produce a range of escaping material from a gravity-dominated boulder population (1--100 km) \citep{leinhardt12} to $\sim$millimeter- to centimeter-size dust spherules formed from vapor condensates \citep{johnson12a}. The infrared flux evolution of an impact-produced dust clump depends on the detailed properties of the generated debris (e.g., \citealt{su19}) and its interaction with the existing background population of planetesimals.  Such background planetesimals exist in HD 166191 as they produce substantial infrared excess ($f_d \sim$0.1) in the quiescent state. The earliest IRAC 3--5 $\mu$m measurements of the system were obtained in 2006 (BMJD 54005) by the {\it Spitzer}/GLIMPSE Survey \citep{glimpse03,glimpse09} as noted by \citet{schneider13}. The IRAC band 1 and 2 fluxes of the system, adopted from the GLIMPSE source catalog \citep{glimpse_cat}, are consistent with the ones from our warm {\it Spitzer} measurements in 2015--2017 within 5\%, suggesting the quiescent state has persisted over a decade. 

As noted in Section \ref{sec:star}, the quiescent SED can be described by two different dust temperatures: $\sim$760 K and $\sim$175 K ($f_d$ of $\sim$4$\times10^{-2}$ and  $\sim$6$\times10^{-2}$, respectively) based on the SED analysis from \citet{schneider13}. The color temperature inferred from the two short IRAC wavelengths is lower ($\sim$650 K), suggesting that the color temperatures derived from two close wavelengths are not representative of the dust temperatures and should only be taken in a relative sense. Assuming blackbody-like grains under optically thin conditions, the SED dust temperatures imply  stellocentric distances of $\sim$0.27 au and $\sim$5 au as the dominant debris locations in the quiescent state. The inferred distance would be much larger if the emission is dominated by small grains. For example, submicron  silicate-like grains, known to be present during the quiescent state as inferred from the prominent 10 $\mu$m feature, can reach $\sim$750 K at a stellocentric distance of $\sim$0.6 au. As we explore the implications of the {\it Spitzer} data, the discussion only applies to the inner $\sim$1 au zone. 

The large-scale infrared brightening obviously points to a huge increase in the debris emission, freshly generated in the inner region during the active state. 
The increase in the observed color temperatures between the two states can be explained by two possibilities: (1) the newly produced impact debris is located closer to the star than the background population in the inner zone, and (2) the dominant grains that produce the 3--5 $\mu$m emission become smaller (i.e., hotter) than the ones in the quiescent state. For the first case, the newly produced debris in the active state needs to be closer to the star by $\sim$30\% to account for the temperature increase (dust temperature $T_d \sim r^{-0.5}$ where $r$ is the distance from the star). For the latter case of accounting for the temperature increase, the dominant grains in the active state need to be $\sim$30\% of the sizes in the quiescence stage  ($T_d \sim a^{-1/6}$, where $a$ is the grain radius assuming an interstellar-median-like composition). Both cases are consistent with the proposed impact scenario. The impact-produced debris is expected to spread to a range of distances from the impact location, and the width of the spread depends sensitively on the impact location: the farther away the impact, the larger the spread of the impact debris \citep{watt21}. In this inner 1 au region, roughly half of the impact debris is found interior to the impact location (see Figure 9 from \citealt{watt21}). Secondly, it is expected that the impact-produced debris would go through collisional cascades particularly in a dense dust clump form, i.e., producing smaller grains consistent with the observed temperature increase. As noted earlier, the observed color temperature also drops during the eclipse event, which is consistent with less heating of the dust due to obscuration.

\begin{figure*}
    \centering
    \includegraphics[width=\linewidth]{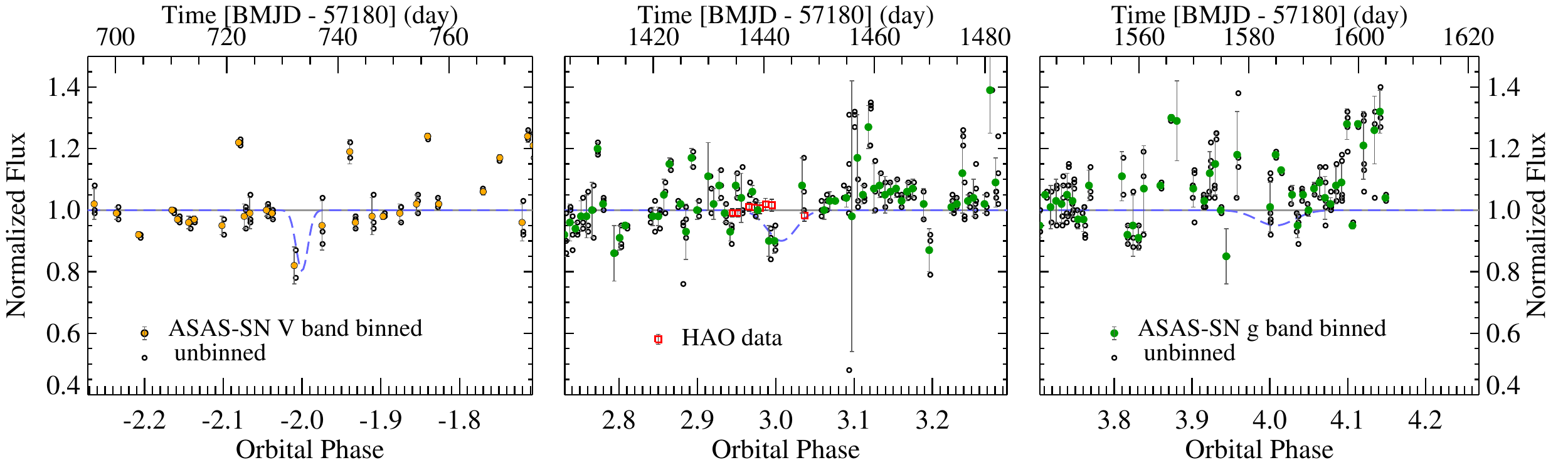}
    \caption{Tentative dust clump evolution as displayed in the optical light curve with the potential curtain models shown in blue dashed line to guide the eyes. The symbols used are the same as in Figure \ref{fig:opt_irac_transit}. The dust clump started as compact, became thicker and larger after $\sim$2 orbital periods, and dissipated (not transiting star) after $\sim$3--6 orbital evolution (details see Section \ref{sec:clump_evo}).}
    \label{fig:clumpevolution}
\end{figure*}

\subsection{Clump Properties and Evolution}
\label{sec:clump_evo}

Based on the curtain model, the obscuring object is likely a dust clump at a semi-major axis of 0.62 au (assuming a circular orbit) given the apparent orbital period of 142$\pm$0.3 days around a 1.6 $M_{\sun}$ star. At that location and assuming a circular orbit, the Keplerian velocity is 47.8 km~s$^{-1}$ (or 5.9 $R_{\sun} \text{day}^{-1}$), suggesting that the star has a stellar radius of 2.37 $R_{\sun}$ using the best derived velocity of 2.5 $R_{\ast} \text{day}^{-1}$. This radius is 18\% larger than what we derived in Section \ref{sec:star}, which already takes the stellar youth into account. This apparent discrepancy can be easily solved if the clump is on an eccentric orbit. 

The transit depth is an estimate of the fraction of the stellar disk that was obscured during the deepest eclipse assuming the clump is optically thick at the observed wavelength. Under the assumption of a circular orbit, it has a comparable size to the star ($\sim$4.1 $R_{\sun}$ or 0.019 au) vertically given the deepest transit depth of 0.88 at $V$ band. Horizontally, it is $\sim$1.5--2.5 times of the stellar diameter ($\sim$0.033--0.055 au) given the best-fit curtain widths. The dust clump is large -- the projected surface area is 1--2$\times10^{23}$ cm$^2$ (i.e., a coverage factor of $\sim$1$\times10^{-4}$ at 0.62 au). Even so, this surface area is a lower limit because it only applies to the part of the cloud that passed in front of the star. We derive below (Section \ref{sec:implication}) a total dust cross section two orders of magnitude larger, based on the fraction of the light from the star that had to be absorbed to account for the increase in the output of the dust at the peak of the outburst in 2019. The apparent discrepancy implies that the clump is likely much larger and on an orbit that is slightly inclined relative to us so the dips were due to structures toward the edge of this cloud. There are, of course, multiple alternative explanations as we further discussed in Section \ref{sec:implication}.

Using the multiwavelength data for dip \#2, we can put some constraints on the dominant grain size in the clump if the occulting object has a uniform column density along the line of sight. In the optically thin case, the optical depth ratio reflects the extinction; the observed wavelength-dependent optical depth ratio in the infrared is consistent with the interstellar extinction curve \citep{gordon21}. The observed range of absorption  ($\sim$0.2--0.4) between $V$ and 3.6 $\mu$m suggests the dominant grain size is less than $\sim$1 $\mu$m for the majority of grain compositions (Figure \ref{fig:qratio}). Intuitively, this size appears to contradict the observed symmetric transit profiles because the orbits of these small grains should be significantly altered by the radiation pressure, i.e., exhibiting elongated trailing/egress profiles  \citep{lecavelier99a}. We will discuss this further in Section \ref{sec:implication}. 
Alternatively, the optical depth ratio could reflect the spatial distribution of the column density if the grains in the clump are large (i.e., $Q_{3.6}/Q_V \sim$1) and experience negligible radiation pressure effects.  In this case, the clump is symmetric along the direction of motion, i.e., thicker at the center and thinner outside, but the thickest part of the clump observed in the infrared is slightly trailing (behind) the thickest part of the clump observed in the optical by $\sim$0.6 days, suggesting density inhomogeneity if optically thin.

It is also likely that the clump is optically thick at $V$ but thin in the infrared wavelengths, i.e., the extinction ratio no longer traces the absorption accurately. In this case, the timing offset between the optical and the infrared wavelength might be a manifestation of grain properties and/or optical depth effects (given the possibility  that the clump is highly structured and asymmetrical along the direction of motion). 

With the best-fit clump parameters, we searched for the potential signs of the clump before and after dips \#1 and \#2 in the optical data assuming a period of $\sim$142 days. There were some tentative shallow dips before and after; however, none of them are significant (more than 3$\sigma$) given the noisy nature of the stellar output (as shown in Figure \ref{fig:clumpevolution}) so we did not attempt any fits to these shallow dips. The most likely detection of a dip is at $\sim$284 days before dip \#1 (at an orbital phase of $-$2) where one single binned data point is $\sim$20\% (2$\sigma$) low. Taken at face value, this tentative dip is consistent with the curtain width of  dip \#1, but $\sim$4 times shallower.  Similarly, there appears to be a very broad and shallower ($\sim$10\%, 1$\sigma$) transit signal $\sim$284 days after dip \#2 (at an orbital phase of 3). Nonetheless, the reality of this dip is challenged by the nightly data taken at the HAO over the week of the expected transit (see the middle panel of Figure \ref{fig:clumpevolution}). Finally, there is no sign of dips at the orbital phase of 4 ($\sim$426 days after dip \#2). As shown in Figure \ref{fig:opt_irac_transit}, there appeared to be two additional dips (in the range of 30--40\% levels) before dip \#2 that were not present prior to dip \#1, but might be present before the tentative, broad, and shallow dip at the orbital phase of 3. It is difficult to determine whether these dips were associated with the main clump or not. If they were, the behavior is consistent with rapid evolution. We note that radiation pressure might also play a role in the early clump evolution because of its short timescale (on the order of an orbital period for grains smaller than the blowout size under a gas-free condition). If so, we would expect a rapid flux increase (due to new grains generated in the clump) followed by a rapid flux decrease (due to radiation pressure blowout) in the infrared light curve in the early phase (a similar behavior of the 2015 light curve of ID8; \citealt{su19}), which is not observed. The dust-clearing timescale is expected to be longer under the condition of a small amount of residual gas due to the combination of radiation pressure and gas drag  \citep{takeuchi01,kenyon_najita_bromley2016,krumolz20}. As suggested by \citet{kenyon_najita_bromley2016}, the dust outflow velocity is on the order of 0.01 au~yr$^{-1}$ at 1 au for a disk with a gas mass of 0.001\% of a typical solar nebula. Such low gas mass would be unobservable by optical and near-infrared techniques. Future investigations are needed to further assess the role of residual gas in the dust clump evolution.

Under the premise that all the transit signals were related (i.e., a dust clump), the optical data suggest a rapid evolution in the clump -- totally dissipating after $\sim$3--6 orbits. Overall, the dust clump started compact, grew thicker and larger quickly after $\sim$2 orbital periods, and became so thin and extended that no trace was seen after $\sim$3--6 orbital evolutions. \citet{kennedy17_rzpsc} show that the shearing rate for an unbound spherical clump is relatively fast, so the inference of rapid evolution of the occulting structure is not particularly surprising. The rapid evolution is also consistent with numerical simulations where the clump phase only lasts for less than $\sim$10 orbital evolutions \citep{jackson12,kral15,watt21}.

\subsection{A Catastrophic Collision between Large Asteroids in the Terrestrial Zone}
\label{sec:implication}

The flux increase between the quiescent and active states is about $\sim$470 mJy at 4.5 $\mu$m. At the distance of the star (101.5 pc) and an assumed dust temperature of 650 K, the flux increase suggests a minimal increase in the dust cross section of 0.065 au$^2$ ($\sim$1.5$\times10^{25}$cm$^2 \thickapprox$200 stellar sizes). Simulations indicate a substantial rate of fragment collisions rapidly following the impact disruption of asteroid-sized bodies \citep[e.g.,][]{delloro15}, which would expedite the initiation of an intense collisional cascade. Collisions at a modest fraction of the orbital velocity at the position of this debris can accelerate substantial amounts of debris to the velocities required to occult the star \citep[e.g.,][]{kenyon05,delloro15,hyodo_genda20}. Assuming the infrared excess comes from small grains generated in a collisional cascade with blowout sizes of 0.5--5 $\mu$m and grain density of 3 g~cm$^{-3}$, the increased dust cross section corresponds to a ``minimum mass" of $\sim$1--4$\times 10^{23}$ g. Such a mass is equivalent to totally breaking up an object with a diameter of $\sim$400--600 km assuming a density of 3 g~cm$^{-3}$ (i.e., a Vesta-size object).  Adopting different temperatures (500--800 K) has little effect on the size of the object (320--700 km). 

We stress that this dust cross section and the resulting mass estimates are truly minimal because the emission might not be completely optically thin and the infrared emission is only sensitive to small grains, not to fragments larger than $\sim$millimeter to centimeter sizes. Giant impacts involving protoplanets leave a substantial fraction of the mass of the colliding bodies in large fragments \citep[e.g.,][]{jutzi2010, benavidez2012, emsenhuber2018, gabriel2020}, suggesting that the total mass required to produce the infrared-emitting dust would be much greater than our estimates. As has been suggested previously, a rapid rise in the infrared generally points to a sudden increase of $\sim$millimeter- to centimeter-size spherules likely produced quickly from impact-produced vapor, which nominally only accounts for a few percent of the total mass from a giant impact \citep[e.g.,][]{watt21}. A recent study by \citet{gabriel21} further suggests that the production of vapor from giant impacts is only favored if the impactors are $\gtrsim$1 \% Earth mass. All point to the fact that the colliding bodies are much larger than our estimate.

The symmetric profiles of the dips imply that the grains in the clump experience very little radiation pressure effect initially, suggesting the dominant grain sizes in the clump are large compared to the observed wavelengths and consistent with the expected sizes of vapor condensates (a few 100 microns to centimeters) from violent collisions \citep{johnson12a}. Given the large mass involved in the collision, the impact-produced clump is very likely to be optically thick initially, and the presence of optically thick dust clumps has been inferred in some young, extremely dusty systems \citep{meng12,su19,su20,melis21,powell21_tic400799224}. Inside an optically thick clump, grains smaller than the nominal blowout size generated through collisional cascades could be retained because stellar radiation pressure would not be effective in removing them, unlike in a typical low-density debris disk. Small grains would accumulate at the dense part of the clump and might quickly disperse (i.e., accelerating the expansion of the clump) when the clump experiences enough shear.
 
Using the optical data, we can further put some lower limits on the expansion of the dust clump by assessing the rate of change in the clump width (sensitive to the projected velocity in the horizontal direction) and transit depth (sensitive to the projected velocity in the vertical direction) relative to the clump center. The width change between dips \#1 and \#2 suggests a projected expansion rate of 0.98 $R_{\ast}$ per orbit ($\sim$110--130 m~s$^{-1}$ assuming a stellar radius of 2--2.38 $R_{\sun}$). Assuming the clump is very optically thick between orbital phases of $-$2 and 0 (dip \#1), the change in the transit depth suggests a projected expansion rate of 0.3--0.4 $R_{\ast}$ per orbit (i.e., $\sim$35--55  m~s$^{-1}$ assuming $\tau \sim$0--0.2 at the orbital phase of $-$2). Interestingly, an expansion velocity of a few hundred m~s$^{-1}$ would fit with the escape velocity of a body a few hundred km in size (Vesta has an escape velocity of $\sim$350 m~s$^{-1}$). As shown in Section \ref{sec:clump_evo}, the dust clump has a projected surface area of 1--2$\times10^{23}$ cm$^{-2}$ at dip \#2. The clump's minimum mass is 8$\times10^{20-22}$ g by adopting grains of $\sim$10--1000 $\mu$m size to provide the necessary surface area. Using a zeroth-order estimate, the mass estimate is equivalent to a body of a few hundred kilometers in diameter.  The estimated expansion velocity (lower limits) is consistent with the mass requirement. 

The projected expansion rate suggests that the clump has a vertical half-width of $\sim$1.1 $R_{\ast}$ at dip \#2 (orbital phase of 1), suggesting the clump is less optically thick. Furthermore, the projected clump size would increase by a factor of $\sim$3 between the orbital phases of 1 (dip \#2) and 3 if continuing with the same velocities in both directions. The increased area of the clump would further broaden the clump width and reduce the thickness ($\tau$) at the center, roughly consistent with no strong transit signal at the orbital phase of 3 and farther. If the expansion of the clump had some kind of acceleration, such as a sudden exposure of small grains produced in the clump center, the clump would very likely be totally disrupted at later orbits. Furthermore,  being on an eccentric orbit, as indicated from the stellar radius discrepancy (Section \ref{sec:clump_evo}), would further facilitate its disruption (A. Jackson et al.\ 2022, in preparation).  Overall, the transit profile evolution is consistent with that of an impact-produced clump under Keplerian shear. 

As the clump is being dispersed, small grains generated in the clump center are likely to collide with other debris (i.e., other large fragments and/or background population) in the system, creating a snowball effect and increasing the infrared flux of the system. We might witness this kind of phenomenon right after the orbital phase 2 (between the labels of C and D in Figure \ref{fig:irac_hd166191}). There appears to be big flux jump between the orbital phases of 2 and 3 by $\sim$50\% and 30\% at 3.6 and 4.5 $\mu$m, respectively, and yet before and after the jump the overall disk flux appeared to be relatively flat. Although there was a NEOWISE measurement between the two {\it Spitzer} visibility windows that supported a linear flux increase, the saturation of the NEOWISE data makes such a trend inconclusive. 

\subsection{Short-term modulation in the HD 166191 infrared light curves?}
\label{sec:short-term} 

Violent impacts involving large asteroid-sized bodies are expected to form thick clouds of debris \citep{jackson12,jackson14,watt21}, and the aftermaths of these impacts produce complex short- and long-term infrared variability due to viewing geometry, dynamical, and collisional evolution in the impact-produced fragments \citep{su19}. Such behaviors are well documented in the prototype of extreme systems around a 35 Myr-old solar-like star, ID8 \citep{meng14,su19}. Our impact hypothesis for HD 166191 is very similar to the ID8's behavior in 2014/2015, showing a ramp of infrared brightness due to the expansion of an optically thick cloud of debris as it underwent Keplerian shear. In addition to the flux increase, ID8's 2014/2015 light curves also exhibited short-term variation on timescales of half the orbital period due to the thickness of the clump along the line of sight (i.e., disk ansae). However, we do not see significant evidence for short-term modulations in HD 166191, aside from the transit signals. Because we are proposing that the mechanism that explains the year-long increase in brightness is essentially the same between HD 166191 and ID8, it is relevant to consider why ID8 displays short-term variations in the infrared light curves, but HD 166191 does not.

Firstly, it is important to note the difference in the observation windows and the orbital periods related to the proposed models.  Due to the position on the sky, the {\it Spitzer} visibility windows for ID8 are generally over 200 days in length, whereas for HD 166191 the {\it Spitzer} visibility windows are only $\sim$39 days. In the model described by \citet{su19}, two kinds of bimodal variation, each at half of the genuine orbital period can occur: one due to the viewing geometry (minimum flux at the disk ansae) if the disk midplane is close to edge on, and the other due to the orbital evolution of the impact fragments at the collisional point and anticollisional line \citep{jackson14}. Given the appearance of the transit signals we expect that the orbital plane of planetesimals in HD 166191 is close to edge on, therefore, the time when the clump passes the disk ansae would be $t_0$ (the time the clump passes the center of the star) plus one-fourth and three-fourths of the orbital period. Unfortunately, these times (marked by the dash-dotted lines in Figure \ref{fig:irac_hd166191}) lie in gaps of the {\it Spitzer} observations. For the modulations due to the collision point and anticollision line, it is difficult to predict the observable signal because we do not know the exact positions relative to the disk ansae. However, the minimum average timescale should be one-forth of the orbital period, i.e., $\sim$35 days (almost equal to the length of the visibility window), making the short-term brightness variations more difficult to detect in the HD 166191 system.

Furthermore, the amplitude and timescale of the short-term brightness variations in the optically thick, impact-produced debris are also strongly dependent on how the fragments (both unaltered boulders and vapor condensates) are released post-impact. Specifically, the distribution of the released debris depends sensitively on the orientation of the collision with respect to the orbital plane around the central star, the mass ratio between the two impacting objects, and the impact angle of the collision,  all producing an anisotropic distribution of the resulting debris (e.g. \citealt{watt21}). For a grazing collision, the  debris tends to have a large velocity dispersion, easily spreading over a large range of semi-major distances so that it is unlikely to remain a coherent clump able to produce short-term modulation. Given the variables, it is not surprising that the HD 166191 dust clump showed no short-term modulation.

\subsection{What Triggers the Onset of the Two States?}
\label{sec:trigger}

The minimum increase in the total dust cross section between the quiescent and active states is on the order of 10$^{25}$ cm$^2$ ($\thickapprox$200 stellar sizes), which is 100 times larger than the projected area of the clump (a few stellar sizes). As discussed in Section \ref{sec:clump_evo}, the clump is likely much larger and on an inclined orbit so that only part of the clump eclipses the star. Nevertheless, an estimate on the increased surface area due to Keplerian shearing with the expansion velocity of the clump and that of new small grains generated through collisional cascades within the dense clump (a condition is expected to be mostly localized) are unlikely to account for the full amount of flux increase between the two states. This suggests that multiple large-scale collisions are required and that the initial collision likely creates several large fragments. It is difficult to determine the exact time of the initial collision using the infrared flux alone (because infrared flux is only sensitive to small dust grains). Nonetheless, the initial collision is likely occurred during the quiescent phase because it takes time to generate enough small grains to display flux increase in the infrared. Subsequent collisions either among them or with the background planetesimals result in several dust clumps and only one transits the star. In fact, the dips prior to dip \#2 (near orbital phase of 0.85--0.95 in Figure \ref{fig:opt_irac_transit} as described at the end of Section \ref{sec:clump_evo}) might be caused by partial occultations of other clumps. 

The question remains: What triggered the onset of these two states?
A number of scenarios could have precipitated it. Given its young age (right after the dispersal of gaseous material), the infrared brightening might be signaling the initial assembling of terrestrial planets through multiple giant impacts. In this case, the system's infrared flux might exhibit multiple, large-scale flux increases in the future. Alternatively, it might be triggered by unseen massive bodies in the system, mostly likely planet-mass objects (because of the constancy of the radial velocity of the star ruling out the presence of stellar companions) and/or some global event like the reconfiguration of giant planets causing intense high-velocity collisions in the inner system \citep{carter_stewart20}. Future data, particularly continuous monitoring in both the optical and infrared wavelengths, would yield better insights into this unique system.

\section{Conclusion }
\label{sec:conclusion}

We report on five years of 3--5 $\mu$m photometry with warm {\it Spitzer} that tracks the debris dust emission in the terrestrial zone of the HD 166191 system. We use publicly available optical measurements over the same time span to characterize the stellar activity and show that the typical rms is 5\%--10\%, consistent with the star being a young late F- to early G-type star. Overall, the infrared output of the system at 3.6 and 4.5 $\mu$m has two states: a quiescent phase before 2018 and an active phase afterward. The infrared fluxes gradually increased starting in mid 2018 and reached a plateau by mid 2019, doubling the total excess emission. Because there is no such long-term and large-degree brightening from the star itself in the optical data, we attribute the flux increase in the infrared to a change in circumstellar debris dust. The amount and rapidity of the increase in the infrared flux requires a catastrophic event, such as a collision between two large bodies ($\gtrsim$ 500 km in diameter) occurring in the terrestrial zone.

During the infrared brightening phase, a sudden drop was observed at both 3.6 and 4.5 $\mu$m but with different depths between the two bands. This event was also seen in the optical as a very deep ($\gtrsim$80\%) dip. A similar deep dip was also seen in the optical 142 days earlier (there are no contemporaneous infrared data). A Bayesian analysis using Gaussian profiles in combination with a Gaussian Process regression finds that the two optical dips were different in both the width and amplitude at $\sim$2$\sigma$ levels. Within the limitations of our sampling, the dip profiles are symmetric. 

Symmetric profiles at all wavelengths imply that the transiting material is symmetric along the direction of motion, and the variable nature of the occultations argues that the transiting object is not a single opaque body. We characterized the multiwavelength transit profiles using a one-dimensional translucent curtain model, described by \citet{kennedy17_rzpsc}. The timings of the two optical dips provide a strong constraint on the orbital period of the transiting object, i.e., 142 days, equivalent to a semi-major axis of 0.62 au around the 1.6 $M_{\sun}$ star assuming a circular orbit. Because the curtain width and its velocity are degenerate, we determined the best-fit parameters using the Markov chain Monte Carlo (MCMC) technique by assuming a circular orbit for the object. Comparing the two optical dips, our modeling reveals that the optical depth and the width of the curtain got thicker and wider with time. For the deeper dip, which was also observed in the infrared, both the depth and width were reduced significantly at longer wavelengths. These trends remain the same whether the object blocks the light from the star only or the disk plus the star. The preferred velocity of the curtain under the assumption of a circular orbit suggests a larger stellar size compared to the expected one given the youth of the star -- a discrepancy that can be easily solved if the curtain is on an eccentric orbit. 

Given the derived transit properties (time evolution and wavelength-dependent width and depth) in combination with the year-long infrared brightening, the obscuring object is most likely a dust clump created by a recent large asteroid collision in the terrestrial zone of the HD 166191 system. From the transit depth and width, one can estimate the minimum size of the clump using the deepest dip in the optical data. Assuming a circular orbit, the clump is comparable in size to the star vertically, and $\sim$2--3 times the stellar diameter horizontally. The infrared observations of the same dip show wavelength-dependent extinction, similar to interstellar extinction, suggesting a dominant submicron particle size. However, grains with such small sizes are expected to experience significant orbital alteration due to radiation pressure, which would manifest as an asymmetric transit profile that is not observed in the data. It is likely that the clump is predominantly made of larger particles such as the impact-produced vapor condensates that shield the smaller ones from some of the radiation pressure. That is, the clump is optically thick in the optical, but less so in the infrared. 

We searched for the potential signals of the clump before and after the dips in the optical data, and found tentative shallow dips that might be associated with the period of 142 days, but none of them are significant ($>$3$\sigma$). These constraints suggest that the dust clump was initially compact, and grew thicker and larger quickly after $\sim$2 orbits, but became so thin and extended that no trace was seen after an evolution of $\sim$3--6 orbits. We further put some lower limits on the expansion of the clump by assessing the change rate using the transit width and depth, and find a projected expansion of $\sim$110--130 m~s$^{-1}$ and $\sim$35--55 m~s$^{-1}$ in the horizontal and vertical directions, respectively. An expansion velocity of a few hundred m~s$^{-1}$ would fit with the escape velocity of a body a few hundred km in size such as Vesta. The actual body is likely to be much larger because the data are only sensitive to the projected expansion velocity. The shearing rate for an unbound dust clump, particularly on an eccentric orbit, is relatively fast, consistent with its rapid evolution. 

The minimum increase in the total dust cross section ($\thickapprox$200 stellar sizes) between the quiescent and active states is 100 times larger than the projected area of the clump (a few stellar sizes). The clump size is likely to be underestimated if the clump is on an eccentric and/or inclined orbit and only part of the clump ellipses the star. Given the expansion velocity due to Keplerian shear,  the clump that caused the transits is not solely responsible for the total flux increase in the infrared, and other multiple large-scale collisions are required. It is likely that the initial collision creates several large fragments that subsequently collide either between them or with the background planetesimals,  resulting in several dust clumps and only one of them created the deep transits and others created none or partial occultations. Such a phenomenon is known to exist in our solar system as many asteroid families were created by past break-ups of a much larger body. 

Several scenarios could have precipitated the onset between the quiescent and active states in HD 166191. The system's young age right after the dispersal of its gas-rich protoplanetary disk suggests that we might be witnessing the early giant impact phase in assembling terrestrial planets in the inner zone. Alternatively, this might be triggered by perturbation from nearby, unseen planet-mass objects, creating orbit-crossing collisions in the existing asteroid population, and/or from some global event like planet migration. Long-term monitoring by {\it Spitzer} reveals that a system's infrared output could remain in a quiescent stage over decadal timescales without major changes even when intense and frequent collisions are expected during the terrestrial-planet-forming phase.  The transient nature of extremely dusty systems further illustrates the importance of continuous monitoring. Future observations of this unique system would further shed light into our understanding of terrestrial-planet formation and overall assembling planetary architecture.

\appendix 
\restartappendixnumbering

\section{Warm {\it Spitzer} Photometry for HD 166191}

\begin{deluxetable*}{ccrrrrcrrrr}
\tablewidth{0pc}
\footnotesize
\tablecaption{The IRAC fluxes of the HD 166191 system\label{tab:irac}}
\tablehead{
\colhead{AOR Key}&\colhead{BMJD$_{3.6}$}&\colhead{$F_{3.6}$}&\colhead{$E_{3.6}$}&\colhead{$F_{IRE,3.6}$}&\colhead{$E_{IRE,3.6}$}  &\colhead{BMJD$_{4.5}$}& \colhead{$F_{4.5}$}&\colhead{$E_{4.5}$}&\colhead{$F_{IRE,4.5}$}&\colhead{$E_{IRE,4.5}$}    \\ 
\colhead{  }&\colhead{(day) }&\colhead{(mJy)}&\colhead{(mJy)}&\colhead{(mJy)}&\colhead{(mJy)}&\colhead{(day)}&\colhead{(mJy)}&\colhead{(mJy)}&\colhead{(mJy)}&\colhead{(mJy)}
}
\startdata
    53434624  &    57187.57434  &     724.088   &      2.246  &     240.688   &      7.591 &    57187.57280   &    740.062   &      2.679  &     424.962   &      5.433     \\ 		
    53434112  &    57191.57553  &     716.637   &      2.324  &     233.237   &      7.614 &    57191.57400   &    745.061   &      1.931  &     429.961   &      5.106     \\   	
    53433600  &    57193.13801  &     709.456   &      3.912  &     226.056   &      8.239 &    57193.13648   &    746.132   &      1.838  &     431.032   &      5.071     \\   	
    53433088  &    57197.12839  &     724.561   &      3.944  &     241.161   &      8.254 &    57197.12687   &    738.824   &      4.089  &     423.724   &      6.250     \\   	
    53432576  &    57200.11510  &     713.706   &      5.190  &     230.306   &      8.917 &    57200.11359   &    743.380   &      3.102  &     428.280   &      5.654     \\   	
    53432064  &    57202.92367  &     717.511   &      5.099  &     234.111   &      8.865 &    57202.92216   &    749.307   &      2.160  &     434.207   &      5.196     
\enddata
\tablecomments{$F$ and $E$ are the flux and uncertainty including the star, while $F_{IRE}$ and $E_{IRE}$ are the excess quantities excluding the star.  This table is published in its entirety in the
machine-readable format. A portion is shown here for guidance regarding its form and content.}
\end{deluxetable*}

Table \ref{tab:irac} shows the 3.6 and 4.5 $\mu$m photometry of the HD 166191 system obtained during the {\it Spitzer} warm mission as described in Section \ref{sec:irac_obs}. The excess emission and its uncertainty were derived by the subtraction of the expected photospheric value (483 and 315 mJy at the 3.6 and 4.5 $\mu$m bands, respectively) with a typical uncertainty of 1.5\% of that value added in quadrature.

\facilities{{\it Spitzer} (IRAC)}

\software{astropy \citep{astropy2013}, 
          \texttt{matplotlib} \citep{Hunter2007matplotlib},
          \texttt{numpy} \citep{vanderWalt2011numpy},
          \texttt{scipy} \citep{virtanen2020scipy},
          \texttt{pymc3}, \citep{salvatier2016pymc3},
          \texttt{theano}, \citep{theano2016theano},
          \texttt{celerite}
\citep{foreman-mackey2017fastScalableGPs,foreman-mackey2018celerite}
}

\acknowledgments 

The authors thank Bruce Gary who obtained daily optical data in May 2019 at HAO, leading to better constraints on the rapid evolution of the dust clump. K.Y.L.S. also thanks Phil Carter and Lewis Watt for their comments on the early version of the draft. We also thank the referee for providing constructive comments that improved the clarity of this manuscript. This work has been supported by NASA ADAP programs (grant No. NNX17AF03G and 80NSSC20K1002). G.M.K. is supported by the Royal
Society as a Royal Society University Research Fellow. The paper is based on observations made with the Spitzer Space Telescope, which was operated by the Jet Propulsion Laboratory, California Institute of Technology. This work has made use of data from the European Space Agency (ESA) mission{\it Gaia} (\url{https://www.cosmos.esa.int/gaia}), processed by the {\it Gaia} Data Processing and Analysis Consortium (DPAC, \url{https://www.cosmos.esa.int/web/gaia/dpac/consortium}). Funding for the DPAC has been provided by national institutions, in particular the institutions participating in the {\it Gaia} Multilateral Agreement. We have used photometry from the ASAS-SN project, for which we are grateful.

\bibliography{ksuref}

\end{document}